\documentclass{article}
\usepackage[utf8]{inputenc}
\usepackage{amsmath,amsfonts}
\usepackage{algorithmic}
\usepackage{algorithm}
\usepackage{array}
\usepackage{textcomp}
\usepackage{stfloats}
\usepackage{url}
\usepackage{verbatim}
\usepackage{graphicx}
\usepackage{multirow}
\usepackage{arydshln}
\usepackage{footnote}
\usepackage{cite}

\title{A Structure-guided Effective and Temporal-lag Connectivity Network for Revealing Brain Disorder Mechanisms}
\author{Zhengwang Xia, Tao Zhou, Saqib Mamoon, Amani Alfakih, Jianfeng Lu}
\date{December 2022}

\begin{document}

\maketitle

\begin{abstract}
Brain network provides important insights for the diagnosis of many brain disorders, and how to effectively model the brain structure has become one of the core issues in the domain of brain imaging analysis. Recently, various computational methods have been proposed to estimate the causal relationship (i.e., effective connectivity) between brain regions. Compared with traditional correlation-based methods, effective connectivity can provide the direction of information flow, which may provide additional information for the diagnosis of brain diseases. However, existing methods either ignore the fact that there is a temporal-lag in the information transmission across brain regions, or simply set the temporal-lag value between all brain regions to a fixed value. To overcome these issues, we design an effective temporal-lag neural network (termed ETLN) to simultaneously infer the causal relationships and the temporal-lag values between brain regions, which can be trained in an end-to-end manner. In addition, we also introduce three mechanisms to better guide the modeling of brain networks. The evaluation results on the Alzheimer's Disease Neuroimaging Initiative (ADNI) database demonstrate the effectiveness of the proposed method. 
\end{abstract}

\section{Introduction}
Functional brain networks have been widely used to understand the principles of brain organization \cite{ji2019mapping, shine2018principles} and explore the sensitive biomarkers of neuropsychiatric diseases \cite{huang2020attention}. Brain networks depict the complex patterns of interactions between brain regions and provide a powerful tool for detecting several brain disorders such as autism spectrum disorder \cite{ronicko2020diagnostic} and Alzheimer's disease \cite{peng2009partial}.

A brain network can be viewed as a collection of nodes and edges, where each node represents a brain region defined by the physiological template, and each edge represents the relationship between two brain regions. Recent years have witnessed an endless stream of research on functional brain network modeling. In general, the relationships represented by edges in these modeling methods can be divided into two categories: functional connectivity (FC) and effective connectivity (EC) \cite{friston1994functional}. FC reflects a statistical dependence between functional magnetic resonance imaging (fMRI) signals from distinct brain regions, which assess the direct or indirect interactions between brain regions. Unlike FC, EC is defined as the causal influence one brain region exerts over another, the purpose of which is to determine the direction of information flow. Due to the lack of directional information in FC, brain network models based on FC may yield suboptimal results in identifying abnormal patterns caused by brain diseases. For example, Li et al.~\cite{li2017novel} constructed a novel effective connectivity network for the diagnosis of mild cognitive impairment (MCI), and the experimental results show that the EC-based method achieves a significant improvement over the FC-based method. Chen et al.~\cite{chen2021estimation} designed a message-passing algorithm to estimate the direction of information flow between brain regions, and the result similarly demonstrated that the EC-based method significantly outperforms FC-based methods in detecting disease-related neuroimaging biomarkers. Overall, the results suggest that EC can provide more effective information for discriminating brain disorders than traditional FC.

Recently, various computational methods have been proposed to estimate the direction of information flow between brain regions, which model the problem from different perspectives \cite{bressler2011wiener, frey2005comparison, shimizu2006linear}. However, these methods have their own limitations and they cannot accurately infer the direction of information flow between brain regions in some cases \cite{smith2011network}. For example, Granger causality (GC), one of the most popular methods for inferring causality between brain regions, assumes that the upstream signal will be repeated by the downstream signal with a certain temporal-lag. However, in practice, this assumption is often invalid in fMRI due to the presence of noise during data acquisition \cite{bressler2011wiener}. Bayesian network (BN), a widely used algorithm for constructing brain effective connectivity network (ECN), restricts the candidate graph structure to be a directed acyclic graph (DAG). However, accumulating studies show that functional interactions in the brain are not acyclic due to reciprocal polysynaptic connections \cite{friston2011functional, markov2014weighted}. Linear non-Gaussian acyclic model (LiNGAM), a data-driven approach to infer the direction of information flow between brain regions from fMRI data, is based on independent component analysis (ICA) to search for solutions. However, ICA requires acquiring a large number of time series data, and the performance of LiNGAM is often not ideal when the fMRI data sample is small \cite{shimizu2006linear}. Based on the above analysis, it can be seen that the previous modeling methods may only be suitable for some specific situations, while several strong assumptions cannot be guaranteed to hold.

In addition to the shortcomings mentioned above, many existing effective connectivity estimation methods ignore a common problem: the temporal-lag value of the information flow between different brain regions should be different. For instance, GC method considers the influence of temporal-lag when inferring the causality between brain regions, but it sets the temporal-lag value to a fixed value, equivalent to the time of a complete brain sampling \cite{seth2015granger}. BN is based on the Bayesian scoring metric to search for the potential optimal graph structure, and the time delay of information transmission is not considered in the search process~\cite{ramsey2017million}. LiNGAM assumes that the time series signal of each brain region is a linear combination of all other brain regions, and there is no time lag between them \cite{shimizu2006linear}. Obviously, none of these assumptions correspond to the actual situation. For example, numerous studies have found highly reproducible temporal-lag patterns in the blood oxygen level-dependent (BOLD) signal of healthy subjects \cite{mitra2014lag, bandt2019connectivity}. Raatikainen et al. have proved that there are large variations in temporal-lag of information propagation within the brain \cite{raatikainen2020dynamic}.

\begin{figure*}[htb]
\centering
\includegraphics[width=0.96\textwidth]{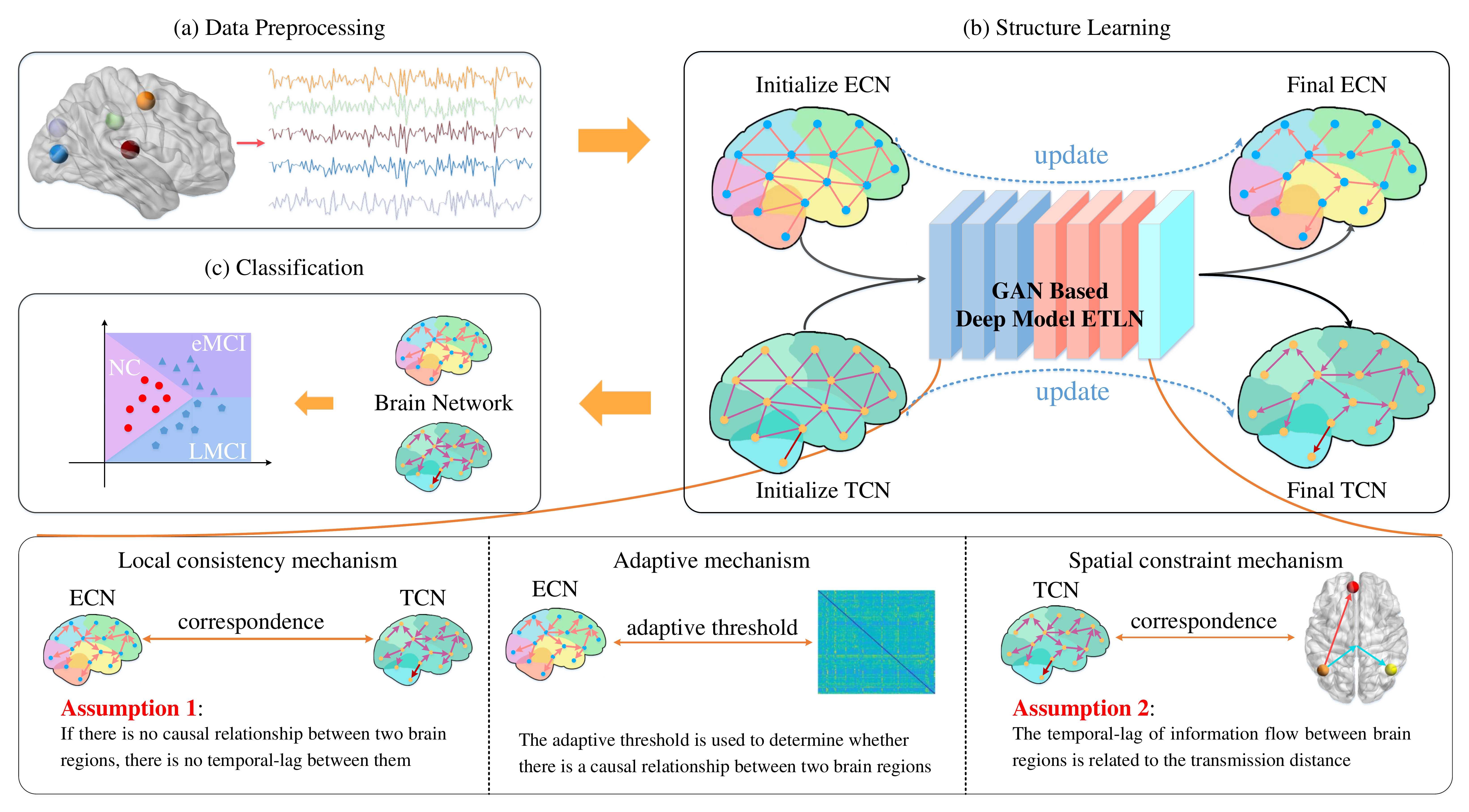}
\caption{Illustration of the proposed framework for brain disorder identification. The framework mainly consists of three steps, including data preprocessing, structure learning, and classification. The data preprocessing step is used to extract the signal values of each brain region from fMRI images. The structure learning step is adopted to estimate causal effects and temporal-lag values between brain regions. In order to better model the interaction between brain regions, we introduce three mechanisms to guide the modeling of brain networks. The classification step is utilized to obtain the final decision result.}
\label{framework}
\end{figure*}

To overcome the shortcomings of traditional ECN estimation methods, we propose a novel deep learning model, named effective temporal-lag neural network (ETLN), to simultaneously infer the direction (effective connectivity network, ECN) and the temporal-lag value (temporal-lag connectivity network, TCN) of information flow between brain regions. Different from existing methods, our method does not impose unrealistic constraints on the underlying graph structure. Our framework can consider the influence of temporal-lag while inferring the effective connectivity structure, while several methods often ignore this point. Figure \ref{framework} illustrates a schematic diagram of the proposed framework. Specifically, we first extract the blood oxygen signal of each brain region from fMRI. Then, data extracted from fMRI are fed into the deep learning model ETLN to estimate the causality and temporal-lag value between brain regions. Finally, a classifier is trained based on the constructed brain network to realize the recognition of abnormal patterns of brain activity. More importantly, we introduce three new mechanisms to obtain more precise estimates of causal and temporal-lag effects between brain regions. For the relationship between ECN and TCN, we introduce the local consistency mechanism to ensure the correspondence between the two candidate graphs. The core idea of the local consistency mechanism is that if there is no causality between the two brain regions in ECN, their corresponding temporal-lag value in TCN should be 0. In order to determine whether there is a causal relationship between the two brain regions in the candidate ECN, we introduce the adaptive mechanism. Furthermore, we have an additional constraint on the data distribution of candidate TCN, that is, the longer the transmission distance between two brain regions, the larger the temporal-lag value of the information flow between them.

The main contributions of this paper can be summarized as follows. First, we propose a new strategy to simultaneously estimate the direction and temporal-lag values of information flow between brain regions via a deep learning pipeline. Second, our brain network modeling method can characterize the nonlinear interaction between brain regions, rather than the traditional linear interaction. Third, we introduce three new mechanisms into the proposed ETLN to guide the construction of the brain network model, which allows for a more accurate assessment of causal and temporal-lag effects between brain regions. The proposed method is verified on the public database (Alzheimer's Disease Neuroimaging Initiative, ADNI) and achieves promising performance compared with other popular benchmark approaches.

\section{Related Work}
\subsection{Brain Network Estimation Methods}

Recently, many studies have found that neurological disorders are associated with abnormal functional integration between some brain regions \cite{ronicko2020diagnostic, huang2020attention}. Over the past few years, several brain network modeling algorithms have been developed for the diagnosis of neurological diseases \cite{zhang2017inter, qiao2016estimating, li2017novel, wang2020large}. These typical brain network construction methods can be divided into two categories from the meaning of functional connection, including 1) correlation-based methods, and 2) causality-based methods.

In the domain of brain network analysis, Pearson correlation is the most popular measure for constructing brain networks, which is widely used to define the strength of functional connectivity, i.e., the temporal correlation between signals from different brain regions. While many meaningful discoveries have been obtained based on correlation brain networks, it is undeniable that the correlation brain network model has its inherent limitations. First, correlation-based brain networks are often too dense to clearly reveal which functional connections are most relevant to brain disorder \cite{wee2014group}. Second, compared with EC, FC lacks the direction of information flow. It may yield suboptimal results for brain disorder identification if ignoring the direction of information flow \cite{bielczyk2019disentangling}.

Besides, methods to construct effective connectivity networks can be divided into two categories: multivariate methods with constraints and multivariate methods without constraints. Bayesian network (BN) is a typical constraint-type method, which has an underlying assumption that the latent graph structure must be a directed acyclic graph \cite{friston2011functional}. However, recent research shows that functional connections between brain regions are not acyclic due to reciprocal polysynaptic connections \cite{markov2014weighted}. It seems that those methods without constraints are superior to those constraint-type methods. But, those methods without constraints almost ignore the temporal-lag value of information flow \cite{mitra2014lag, bandt2019connectivity}. More details about temporal-lag will be given in the next subsection.

\subsection{Temporal-lag Value Estimation Methods}

In the past few decades, there have been few studies concerning the estimation of information propagation lag between brain regions. Granger causality (GC) is a popular effective connectivity estimation method to measure the causal effect of one brain region on another brain region \cite{bressler2011wiener}. It is based on the idea that the cause of an event cannot come after its consequence, which is one of the causal discovery methods that takes into account the influence of temporal-lag. Unfortunately, the temporal-lag value of GC is set to a fixed value when searching the causal relationship between brain regions \cite{seth2015granger}. However, many scholars have found that there is lag variability among different brain regions, that is, the temporal-lag values are not consistent between different brain regions, and this phenomenon has been demonstrated by many studies\cite{ringo1994time, de2010temporal}. In addition to GC, there is another class of studies that define the temporal-lag value across brain regions as a cross-correlation function between two BOLD signals \cite{mitra2014lag, mitra2015lag}. This type of approach also has its own drawbacks. The first controversial point is that the definition of temporal-lag may be too simple and does not correspond to the real situation. The second point is that this method separates the calculation of temporal-lag value from the identification of causality. The temporal-lag should be zero when there is no causality between two brain regions, which is not considered in the definition of such methods.

\section{Data Acquisition and Processing}

In this study, we used a total of 149 subjects' resting state fMRI data from the publicly available Alzheimer's Disease Neuroimaging Intiative (ADNI) database\footnote{http://adni.loni.ucla.edu}, which includes normal controls (NC), early MCI (eMCI), and late MCI (LMCI). Notably, some subjects in the ADNI dataset were recruited at regular intervals, which lead to the existence of different subjects belonging to the same subject. As discussed \cite{wen2020convolutional}, such case can bring the data leakage issue. Because the samples of the same subject collected at different times could be divided into the training set and test set, respectively, which brings an unfair evaluation. Therefore, in order to make the subsequent evaluation results more convincing, we deduplicate the dataset to ensure that there are no two samples from the same subject. The demographic information can be listed in Table \ref{dataInfo}.

In this dataset, each subject signed the written informed consent form after a full written and verbal explanation. This research was approved by the Research Ethics Board of ADNI\footnote{http://adni.loni.usc.edu/study-design/ongoing-investigations/}. All subjects were scanned with the same protocol using 3.0T Philips Achieva scanners. The scanning parameters are as follows: repetition time (TR) = 3000 ms, echo time (TE) = 30 ms, flip angle = $80^\circ$, imaging matrix = $64 \times 64$, slices = 48 and slice thickness = 3.3 mm. For fMRI data, we apply the standard procedures as follows. First, the first 5 volumes of each subject were discarded before preprocessing to avoid noise signals, and then the remaining 135 volumes were reserved for the subsequent analysis. All the functional images were registered to the first image and transformed into the Montreal Neurological Institute (MNI) space with a resample voxel size of $3\times3\times3$ mm$^3$. Subsequently, Conn Toolbox 20b\footnote{https://web.conn-toolbox.org/}, a Statistical Parametric Mapping (SPM12) based preprocessing pipeline, was used to perform outlier detection, direction segmentation and normalization, linear detrending, and functional smoothing with a Gaussian kernel of 8mm full width half maximum (FWHM), etc. Finally, the time series of each brain region is extracted from the preprocessed images based on the AAL atlas.

\begin{table*}[!ht]
\caption{\label{dataInfo} Demographic information of the used dataset.}
\centering
\begin{tabular}{llll}
       \hline
            Group & NC & eMCI & LMCI\\
       \hline
            Male/Female & 22/29 & 21/37 & 24/16\\
            Age(mean$\pm$STD) & 75.2$\pm$6.9 & 72.2$\pm$6.9 & 72.4$\pm$8.0 \\
        \hline
    \end{tabular}
\end{table*}

\section{Methodology}

\par Previous fMRI studies for inferring causality in brain regions suffer from two shortcomings. First, existing methods often impose certain restrictive assumptions on candidate graphs when inferring the causal relationship between brain regions, and the plausibility of these methods has been questioned \cite{friston2011functional}. Second, most methods ignore the influence of temporal-lag, or simply set the temporal-lag value of all brain regions to a fixed value \cite{seth2015granger}. To overcome the two issues, we propose a novel GAN-based neural network, named effective temporal-lag network (ETLN), to simultaneously learn causality and temporal-lag values between brain regions. We provide the details of the proposed ETLN below.

\subsection{Overview of ETLN}
\begin{figure*}[htb]
\centering
\includegraphics[width=0.9\textwidth]{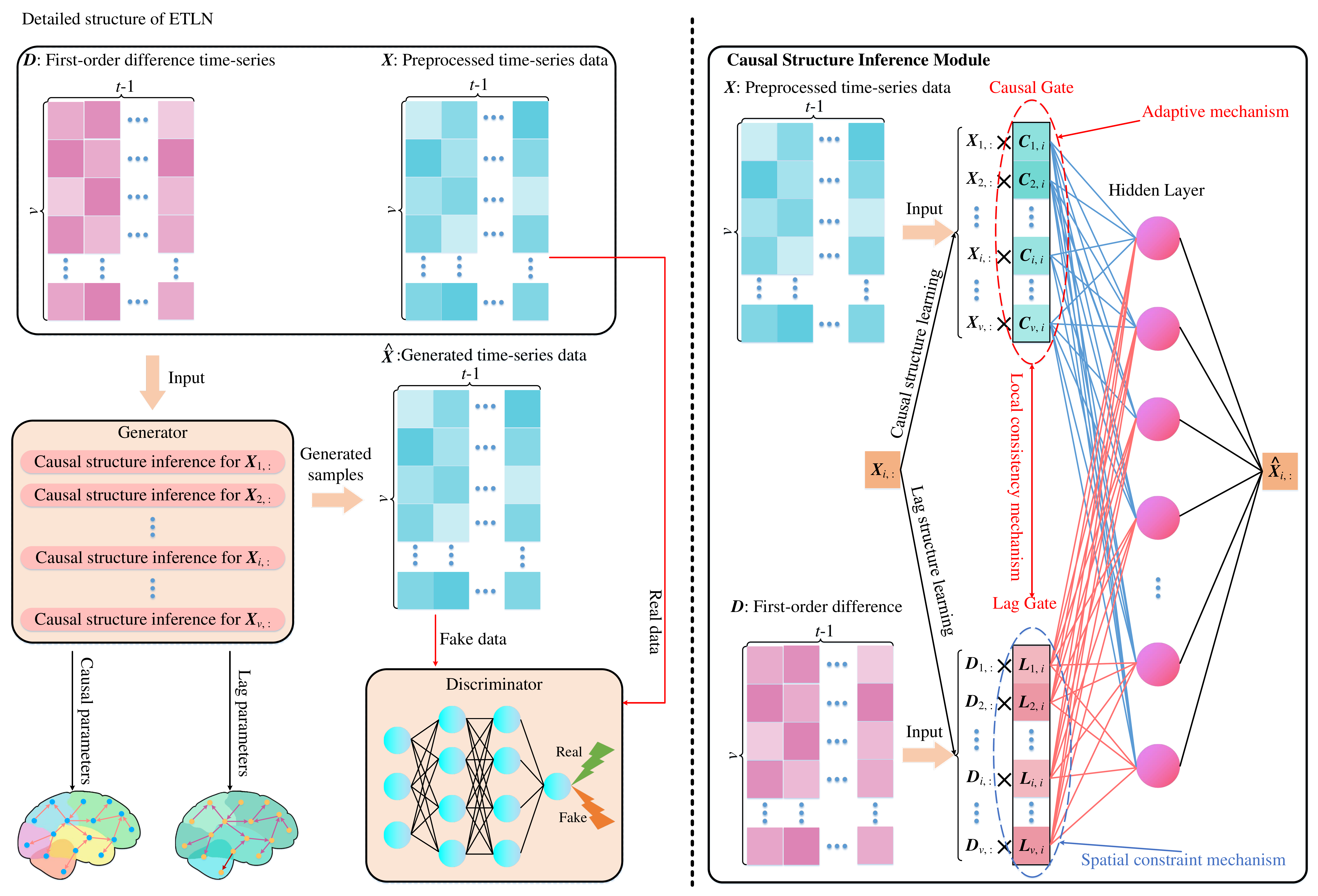}
\caption{The detailed structure of ETLN. The time series data $\textbf{\emph{X}}$ and its first-order difference $\textbf{\emph{D}}$ of each subject are fed into the generator network to estimate causal effects and temporal-lag values between \emph{v} brain regions. Based on the two inputs, the generator network will generate fake time-series data $\hat{\textbf{\emph{X}}}$. The objective of the discriminator network is to distinguish whether the input samples are from real data $\textbf{\emph{X}}$ or fake data $\hat{\textbf{\emph{X}}}$.}
\label{ETLN}
\end{figure*}

The detailed structure of ETLN is shown in Fig. \ref{ETLN}, which embeds the causal relationship and temporal-lag values into a generator network as parameters to be optimized. To clearly describe the proposed model, we first give notations as follows. Let $\textbf{\emph{X}} \in\mathbb{R}^{v \times t}$ denote the obtained time series data from preprocessed fMRI images, where \emph{v} and \emph{t} represent the number of brain regions and the length of the time series, respectively. $\textbf{\emph{D}} \in\mathbb{R}^{ v \times (t-1)}$ is the first-order differences of the time series data \textbf{\emph{X}}. Notably, \textbf{\emph{X}} and \textbf{\emph{D}} are jointly used as the input of the generator network for training, and the dimensions of both are $v \times (t-1)$. The reason why the dimension of \textbf{\emph{X}} is not $v \times t$ will be given in subsection \ref{struc_inference}. Based on the two inputs, the generator network will generate fake time-series data $\hat{\textbf{\emph{X}}}$ that can match the real time-series data \textbf{\emph{X}} as closely as possible. The discriminator network takes real fMRI time series data \textbf{\emph{X}} and fake time series data $\hat{\textbf{\emph{X}}}$ as inputs and tries to find a mapping that can distinguish them. The generator network is composed of \emph{v} causal structure inference modules, where each module is designed to search for the direct causes of the corresponding brain region and the temporal-lag values from these causes to the brain region. In the causal structure inference module, we embed two gates, namely the causal gate $\textbf{\emph{C}}_{:, i} \in \mathbb{R}^{v \times 1}$ and lag gate $\textbf{\emph{L}}_{:, i} \in \mathbb{R}^{v \times 1}$, to preserve the causal effects and temporal-lag values from other brain regions to the $i$-th brain region. After the model is well-trained, the causal relationship and temporal-lag values between \emph{v} brain regions can be obtained from the parameters of the generator network. By splicing the causal parameters and lag parameters of all brain regions, two matrices with dimensions equal to $v \times v$ can be obtained, which is the solution we have been looking for, namely the effective connectivity network $\textbf{\emph{C}} \in \mathbb{R}^{v \times v}$ and the temporal-lag connectivity network $\textbf{\emph{L}} \in \mathbb{R}^{v \times v}$.

\subsection{Formulation of Structure Inference}
\label{struc_inference}

\par The current mainstream methods are based on the idea of multivariate regression to estimate the causal relationship between variables \cite{carballedo2011functional, kalainathan2018sam}. The main advantage of these methods is that they are data-driven and do not take any restrictive assumptions about the structure of the causal graph. Such methods can retrieve excitatory, inhibitory, and bidirectional connections between variables. The core idea behind them is that each univariate component $\textbf{\emph{X}}_{i, :}$ is a mixture of the remaining components $\textbf{\emph{X}}_{j, :}, j \neq i$, and their general form can be defined by
\begin{equation}\label{eq1}
    \underset{\textbf{\emph{C}}_{:, i}}{\operatorname{ \arg\,min}} \parallel\textbf{\emph{X}}_{i, :} - \textbf{\emph{C}}_{:, i}^{T} \textbf{\emph{X}}\parallel_{2}^{2},
\end{equation}
where $\textbf{\emph{X}} \in \mathbb{R}^{v \times t}$ is the time series data extracted from fMRI, and $\textbf{\emph{X}}_{i, :}$ denotes the $i$-th row of \textbf{\emph{X}}. $\textbf{\emph{C}}_{:, i} = [\textbf{\emph{C}}_{1, i}, \dots , \textbf{\emph{C}}_{v, i}] $ is the regression coefficients, which denotes the causal effect of all other brain region on the $i$-th brain region. The superscript $^{T}$ denotes the transpose operation. Besides, $\textbf{\emph{C}}_{i, i}$ is set 0 to avoid self-loops. 

However, such methods ignore the effects of the temporal-lag. The transmission of information between two brain regions takes a certain amount of time, even if the spatial distance between the two brain regions is very close. Therefore, considering the influence of the temporal-lag, Eq. (\ref{eq1}) can be reformulated as follows:
\begin{equation}\label{eq2}
    \underset{\textbf{\emph{C}}_{:, i},\; \textbf{\emph{L}}_{:, i}}{\operatorname{ \arg\,min}}  \parallel\textbf{\emph{X}}_{i, :} - \emph{f}_{i}(\textbf{\emph{C}}_{:, i}, \textbf{\emph{X}}, \textbf{\emph{L}}_{:, i}, \textbf{\emph{D}})\parallel_{2}^{2},
\end{equation}
where the output of $\emph{f}_{i}\left(\cdot \right)$ is the predicted signal value of $i$-th brain region. The specific definition form of $\emph{f}_{i}\left(\cdot \right)$ will be given below.
\textbf{\emph{X}} is used to model the causal relationship between brain regions, and matrix \textbf{\emph{D}} is the first-order difference time series of \textbf{\emph{X}}, which is used to model the temporal-lag relationship between brain regions. The reason why \textbf{\emph{D}} can model the temporal-lag relationship between brain regions will be discussed later. Similar to the above-mentioned methods, the core idea behind our approach is that the signal value of each brain region can be jointly generated by the causal and temporal-lag effects of all other brain regions on that brain region. Likewise, $\textbf{\emph{L}}_{i, i}$ is set to 0 to avoid self loops.

\par
\begin{figure}[htb]
\centering
\includegraphics[width=0.5\textwidth]{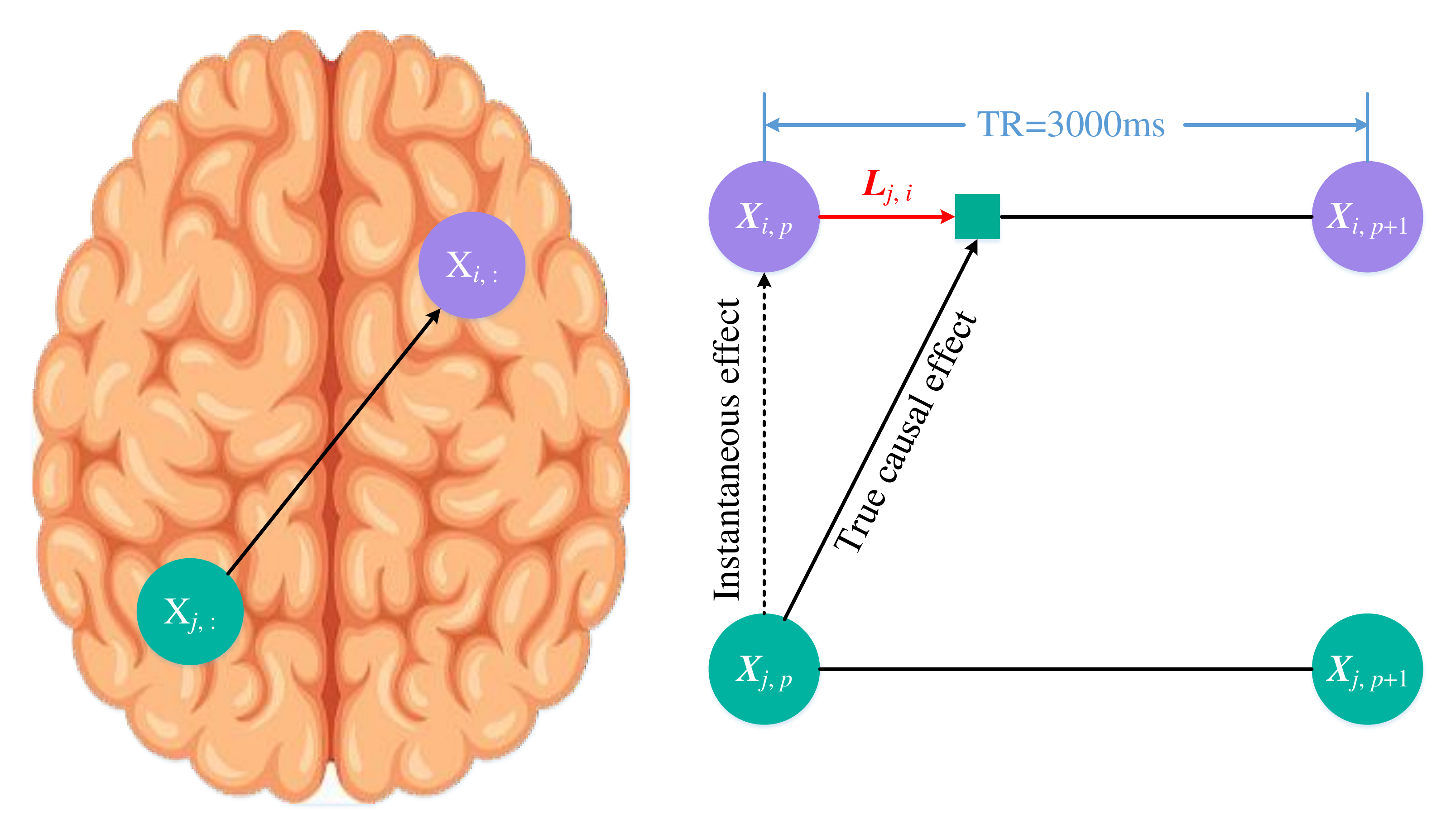}
\caption{Schematic diagram for temporal-lag relationship description.}
\label{temporal-lag}
\end{figure}

 A key question is how to simultaneously model causal and temporal-lag relationships between brain regions, namely the specific form of $\emph{f}_{i}$ in Eq. (\ref{eq2}). For clarity, we show the schematic diagram in Fig.~\ref{temporal-lag}. Suppose $\textbf{\emph{X}}_{i, :}$ and $\textbf{\emph{X}}_{j, :}$ represent the time series data of two brain regions, and there is a causal link from $\textbf{\emph{X}}_{j, :}$ to $\textbf{\emph{X}}_{i, :}$, which can be denoted as $\textbf{\emph{X}}_{j, :} \rightarrow \textbf{\emph{X}}_{i, :}$. A schematic diagram of the information transmission between two brain regions is shown on the right side of Fig. \ref{temporal-lag}. However, these methods often ignore the influence of temporal-lag, and the causal effect between two brain regions is defined as $\textbf{\emph{X}}_{j, p} \rightarrow \textbf{\emph{X}}_{i, p}$. $\textbf{\emph{X}}_{i, p}$ represents the signal value of the $i$-th brain region at the $p$-th time point. This type of methods assumes that the cause and effect can be produced at the same moment, i.e., there is no time delay in the transmission of information between two brain regions, also known as an instantaneous effect. 

Several studies have demonstrated that there is a temporal-lag in the transmission of information across brain regions \cite{raatikainen2020dynamic, raatikainen2017combined}, and the range of temporal-lag is shorter than the measurement rate of fMRI \cite{rajna2015detection}. Therefore, the true causal effect between two brain regions should be defined as $\textbf{\emph{X}}_{j, p} \rightarrow \textbf{\emph{X}}_{i, \left(p + \textbf{\emph{L}}_{j, i}\right)}$, the most crucial point is that the moment when the cause occurs should be before the moment when the effect occurs. $\textbf{\emph{L}}_{j, i}$ denotes the temporal-lag value from the $j$-th brain region to the $i$-th brain region. In addition to instantaneous and true causal effects, we argue that there is also a temporal-lag effect. The temporal-lag effect should be from the instantaneous effect to the true causal effect along the time dimension, as shown by the red directed line in Fig. \ref{temporal-lag}. According to the law of vector addition, we can deduce that the true causal effect should be equal to the sum of the instantaneous effect and the temporal-lag effect. Based on this, we can derive the following equation:
\begin{equation}\label{eq3}
    \textbf{\emph{C}}_{j, i} \cdot \textbf{\emph{X}}_{j, p} = \textbf{\emph{X}}_{i, p} + \textbf{\emph{L}}_{j, i} \cdot \left(\textbf{\emph{X}}_{i, p+1} - \textbf{\emph{X}}_{i, p}\right),
\end{equation}
where the term $\textbf{\emph{C}}_{j, i} \cdot \textbf{\emph{X}}_{j, p}$ means the true causal effect. $\textbf{\emph{X}}_{i, p}$ corresponds to the instantaneous effect, and the term $\textbf{\emph{L}}_{j, i} \cdot \left(\textbf{\emph{X}}_{i, p+1} - \textbf{\emph{X}}_{i, p}\right)$ represents the temporal-lag effect. Generalizing Eq. (\ref{eq3}) to a multivariate case, we can obtain
\begin{equation}\label{eq4}
\begin{split}
    &\hat{\textbf{\emph{X}}} = \textbf{\emph{C}}^{T} \times \textbf{\emph{X}} - \textbf{\emph{L}} \times \textbf{\emph{D}},\\
    &\textbf{\emph{D}} = \frac{100}{TR} \cdot \left(\textbf{\emph{X}}_{:, 2:t} - \textbf{\emph{X}}_{:, 1: t-1}\right),\\
    &\textup{with}~~~~\textbf{\emph{C}}_{i, i} = 0, \textbf{\emph{L}}_{i, i} = 0, i = [1, \dots, v],
\end{split}
\end{equation}
where $\hat{\textbf{\emph{X}}}$ represents the predicted time series data, and \textbf{\emph{D}} is the first-order difference of matrix \textbf{\emph{X}}. $\textbf{\emph{X}}_{:, 2:t}$ denotes all samples of matrix \textbf{\emph{X}} from column 2 to column \emph{t}. The value ranges of causal gate \textbf{\emph{C}} and lag gate \textbf{\emph{L}} are set in the range of [-1, 1] and [0, 1], respectively. It is worth noting that the dimension of matrix \textbf{\emph{D}} is $v \times (t-1)$. Since the causal structure inference module performs pairing training based on the samples of \textbf{\emph{X}} and \textbf{\emph{D}}, thus the dimension of the two needs to be consistent. In this case, the last column of the matrix \textbf{\emph{X}} is discarded, so the dimension of \textbf{\emph{X}} in Fig. \ref{ETLN} is $v \times (t-1)$. Besides, the numerical range of \textbf{\emph{D}} is limited by multiplying by the constant $\frac{100}{TR}$. The reasons for this can be attributed to the following points. First, many studies based on other imaging techniques have reported that the signal transmission delay across brain regions is in the range of 0-100ms \cite{de2010temporal, ringo1994time, smith2011network}, which is much smaller than the measurement time interval of fMRI used in this paper. Second, this design is more conducive to the training of ETLN. Since the causal structure inference module is a two-branch network, if the lag gate is represented by the real numerical range of 0-100, it will lead to the numerical imbalance of the two branches, making the model difficult to train. After setting a $\frac{100}{TR}$ bound for \textbf{\emph{D}}, the lag gate can be set in the range of 0-1, which helps to solve the imbalance problem between the two branches.

\subsection{Constraint Mechanisms}
\label{constraints}
\par In addition to giving a formal definition, we introduce three mechanisms: local consistency mechanism, adaptive mechanism, and spatial constraint mechanism, to better guide the modeling of brain networks. The details of the three mechanisms will be given below.

\par It is worth noting that there is local consistency in the causal and temporal-lag relationships between brain regions. Specifically, for those brain regions with a causal link, the temporal-lag value between them is unclear. But for those brain regions without the causal link, the temporal-lag value between them should be 0. As no information is transferred between two brain regions, which indicates that no time delay between them. Therefore, a local consistency loss is introduced to ensure that the temporal-lag value of those brain regions without a causal link is 0, and its calculation can be defined as follows:
\begin{equation}\label{eq5}
\begin{split}
    &index = \Psi\left(\textbf{\emph{C}}, \mu \right), \\
    &\mathcal{L}_{local} = CE\left(\Vec{\textbf{\emph{L}}}_{index, index}, \Vec{\textbf{\emph{Z}}} \right),\\
\end{split}
\end{equation}
where $\mu$ is a threshold for determining whether there is a causal link between brain regions. The absolute value of the weights in \textbf{\emph{C}} less than $\mu$ is considered as no causal link between brain regions. $\Psi(\cdot)$ returns a set that includes the indices of those brain regions without causal links. $\textbf{\emph{L}}_{index, index}$ is the induced sub-graph of matrix \textbf{\emph{L}}. $\textbf{\emph{Z}}$ is a zero matrix with the same dimension as $\textbf{\emph{L}}_{index, index}$. The symbol \,$\Vec{}$\, means to convert a matrix to a vector, and \emph{CE} refers to the cross-entropy loss.

\par For the threshold $\mu$ in Eq. (\ref{eq5}), it is usually defined in two ways. First, it can be set to a fixed value, and the other is to introduce a dynamic mechanism. Second, the value of $\mu$ is not fixed, and its value varies under different situations. Due to differences among individuals, we utilize the second way to set different values of $\mu$. To achieve this, we introduce an adaptive mechanism to assess whether there is a causal link between brain regions, which is defined by
\begin{equation}\label{eq6}
    \mu = mean(\textbf{\emph{C}}),
\end{equation}
where $mean(\cdot)$ is a function that returns the mean of all elements in matrix \textbf{\emph{C}}. During the optimization process of the network ETLN, the weight of the parameter \textbf{\emph{C}} is not fixed. Therefore, the value of $\mu$ is also constantly changing within the process of network optimization.

\par
\begin{figure}[!t]
\centering
\includegraphics[width=0.4\textwidth]{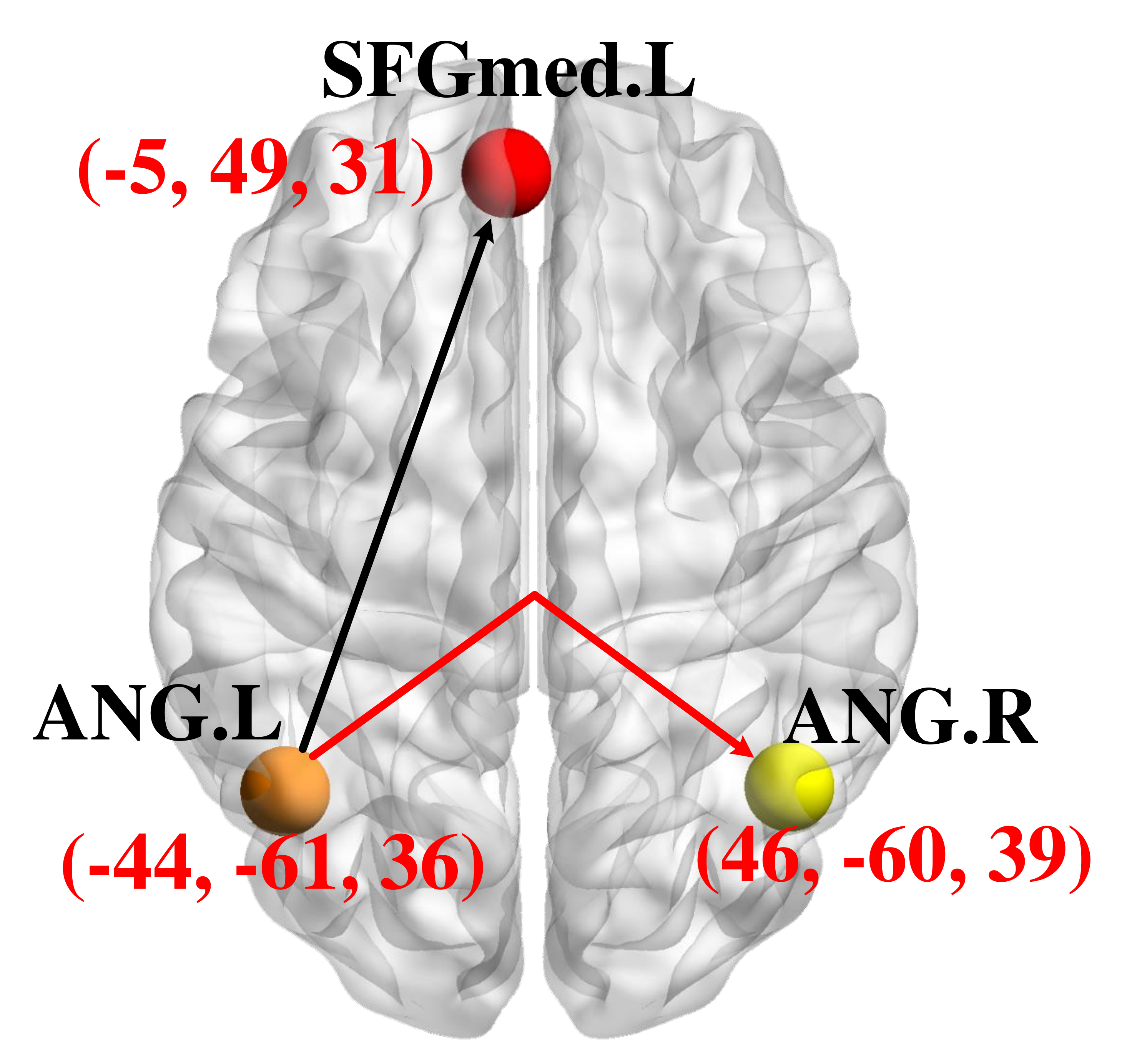}
\caption{Schematic diagram for brain spatial distance.}
\label{spatial-distance}
\end{figure}

 Existing studies have shown that the information transmission delay between the ipsilateral hemispheres is shorter, and the information transmission across hemispheres often requires more time \cite{ringo1994time, meszlenyi2017resting}. In general, the delay of information transmission may have a certain correlation with the transmission distance. Thus, we introduce the spatial distance prior information of the brain to constrain the data distribution of matrix \textbf{\emph{L}}. Then, the first point is how to define the spatial distance prior information of the brain. For the convenience of illustration, we give a schematic diagram as shown in Fig. \ref{spatial-distance} to explain how to define the prior information. In Fig.~\ref{spatial-distance}, each node represents a brain region. Near the node, we give the abbreviation of the brain region and the corresponding center coordinates, respectively. Suppose there is a causal link from ANG.L to SFGmed.L. Since both brain regions are located in the left hemisphere, the transmission distance between them is defined as the distance between the center coordinates of the two brain region, i.e., $len_{ANG.L \rightarrow SFGmed.L} = \sqrt{(-44-(-5))^{2}+(-61-49)^{2}+(36-31)^{2}} = 116.82$. Suppose there is a causal link from ANG.L to ANG.R. Since the two brain regions are located in different hemispheres, the definition of the transmission distance between them is different from the previous one. Studies have shown that the interaction of the left and right hemispheres of the brain relies on the corpus callosum \cite{van2011does}, which is located roughly in the center of the brain. In this study, the coordinate (0, 0, 0) is used to represent the center coordinate of the corpus callosum. Thus, $len_{ANG.L \rightarrow ANG.R} = \sqrt{(-44)^{2}+(-61)^{2}+36^{2}} + \sqrt{46^{2}+(-60)^{2}+39^{2}}= 168.45$. Based on the above calculation method, the pairwise spatial transmission distances between all brain regions of the AAL atlas can be obtained, and the prior information is stored in the matrix \textbf{\emph{P}}. The center coordinates of each brain region in the AAL atlas can be obtained from~\cite{yu2013aberrant}. When the required prior information \textbf{\emph{P}} is obtained, the spatial constraint loss is defined as follows:
\begin{equation}\label{eq7}
    \mathcal{L}_{spatial} = CE\left(\Vec{\textbf{\emph{L}}}_{\overline{index}, \overline{index}}, \Vec{\textbf{\emph{P}}}_{\overline{index}, \overline{index}} \right),
\end{equation}
where $\overline{index}$ is the complement of the set $index$ in Eq. (\ref{eq5}). The aim of the above formula is to constrain the distribution of temporal-lag values for those brain regions with causality.

\subsection{Model Training and Implementation}
\par The overall objective function of ETLN can be derived by combing the local consistency loss $\mathcal{L}_{local}$ and the spatial constraint loss $\mathcal{L}_{spatial}$, as follows:
\begin{equation}\label{eq8}
\begin{split}
    &G_{loss} = MSE\left(\hat{\textbf{\emph{X}}}, \textbf{\emph{X}} \right) + \mathcal{L}_{local} + \mathcal{L}_{spatial}, \\
    &D_{loss} = KL(\textbf{\emph{X}}~||~\hat{\textbf{\emph{X}}}), \\
\end{split}
\end{equation}
where $G_{loss}$ is the generator loss, and $MSE(\cdot)$ is the mean square loss. $D_{loss}$ is the discriminator loss, which minimizes the data distribution difference between the fake data $\hat{\textbf{\emph{X}}}$ and the real data $\textbf{\emph{X}}$ by Kullback-Leibler divergence.

\par The detailed structure of the ETLN network is as follows: each causal structure inference consists of a 1-hidden layer with 200 Tanh units. The discriminator network consists of a 2-hidden layer with 200 hidden LeakyReLU units. The causal and lag gates are initialized to 1, except the self-loop term is set to 0. A standard Adam optimizer with a learning rate of 0.001 is used to optimize the model. The epochs of both generator and discriminator are set to 1000. Our model is implemented in PyTorch using an NVIDIA GeForce 1080Ti GPU with 11 GB memory. In order to obtain more accurate modeling results, the modeling is repeated 32 times for each subject, and the average value is taken as the final result.

\section{Experiments}

In this section, Section \ref{sub_comparsion} first briefly describes the details of the 7 comparison methods. Then, the specific details of the classification step are provided in Section \ref{sub_procedure}. Finally, Sections \ref{sub_classification} and \ref{sub_ablation} present the experimental results of comparative methods and ablation study, respectively.

\subsection{Comparison Methods}
\label{sub_comparsion}
To validate the effectiveness of the proposed method, we compare it with many popular methods. According to whether the brain network contains directional information, these methods can be divided into two types: correlation-based methods (i.e., Pearson correlation-based method (PC) \cite{ronicko2020diagnostic}, sparse representation method (SR) \cite{peng2009partial}, low-rank representation method (LR) \cite{qiao2016estimating} and sparse low-rank representation method (SLR) \cite{qiao2016estimating}) and causality-based methods (i.e., Granger causality based method (GC) \cite{bressler2011wiener}, linear non-Gaussian acyclic model (LiNGAM) \cite{shimizu2006linear} and transfer entropy (TE) \cite{parente2021modelling}). In fact, causality-based methods should also include Bayesian network (BN) \cite{friston2011functional}. However, BN is not suitable for modeling large-scale (more than 20 nodes) data. The optimization time of BN grows exponentially as the number of nodes increases. Since BN is too time-consuming, we did not add it for comparison. We now briefly summarize these competing methods as follows.
\begin{enumerate}
    \item PC method \cite{ronicko2020diagnostic}. In the PC method, the edge weight is defined as the Pearson correlation coefficient between distinct brain regions.
    
    \item SR method \cite{peng2009partial}. This method adds an $L_{1}$ regularization constraint to the graph to exclude confounding effects between brain regions.
    
    \item LR method \cite{qiao2016estimating}. This model introduces modular prior knowledge in brain network construction via a low-rank constraint.
    
    \item SLR method \cite{qiao2016estimating}. This method utilizes both sparse and low-rank constraints to the brain network weight matrix.
    
    \item GC method \cite{bressler2011wiener}. This method determines whether there is a causal link from the $i$-th brain region to the $j$-th brain region based on the Granger causality test. 
    
    \item LiNGAM method \cite{friston2011functional}. The core idea of this method is that the blood oxygen signal data of each brain region is a linear combination of the signals of all other brain regions with no time delay. The regression coefficients are regarded as the edge weights of the brain network.
    
    \item TE method \cite{parente2021modelling}. This method estimates the direction of information flow between brain regions based on the concept of Shannon entropy.
\end{enumerate}
    
\begin{table*}[!ht]
\caption{\label{classification}Performance comparison of different methods.}
\centering
\resizebox{\linewidth}{!}{
\begin{tabular}{c|c|cccc|cccc|cccc}
        \hline\hline
            & \multirow{2}*{Methods} & \multicolumn{4}{c|}{NC vs. eMCI}  & \multicolumn{4}{c|}{NC vs. LMCI} & \multicolumn{4}{c}{eMCI vs. LMCI} \\
        \cline{3-14}
            & & ACC & SEN & SPE & F1 & ACC & SEN & SPE & F1 & ACC & SEN & SPE & F1\\
        \hline
            \multirow{4}*{Correlation based methods} & PC & 68.81 & 66.67 & 70.69 & 66.67 & 70.33 & 76.47 & 62.50 & 74.29 & 65.31 & 79.31 & 45.00 & 73.02 \\
            & SR & 67.89 & 62.75 & 72.41 & 64.65 & 68.13 & 72.55 & 62.50 & 71.84 & 64.29 & 75.86 & 47.50 & 71.54 \\
            & LR & 69.72 & 66.67 & 72.41 & 67.33 & 74.73 & 80.39 & 67.50 & 78.10 & 67.35 & 75.86 & 55.00 & 73.33 \\
            & SLR & 59.63 & 52.94 & 65.52 & 55.10 & 64.84 & 78.43 & 47.50 & 71.43 & 62.24 & 77.59 & 40.00 & 70.87 \\
        \hline
            \multirow{5}*{Causality based methods} & GC & 62.39 & 50.98 & 72.41 & 55.91 & 67.03 & 76.47 & 55.00 & 72.22 & 61.22 & 75.86 & 40.00 & 69.84 \\
            & LiNGAM & 54.13 & 47.06 & 60.34 & 48.98 & 58.24 & 68.63 & 45.00 & 64.81 & 58.16 & 75.86 & 32.50 & 68.22 \\
            & TE & 56.88 & 56.86 & 56.90 & 55.24 & 64.84 & 76.47 & 50.00 & 70.91 & 51.02 & 63.79 & 32.50 & 60.66 \\
            & ETLN$^{E}$ & \textbf{74.31} & \textbf{70.59} & \textbf{77.59} & \textbf{72.00} &  \textbf{78.02} & 82.35 & \textbf{72.50} & \textbf{80.77} & \textbf{73.47} & 82.76 & \textbf{60.00} & \textbf{78.69} \\
            & ETLN$^{T}$ & 71.56 & 64.71 & \textbf{77.59} & 68.04 & 72.53 & \textbf{84.31} & 57.50 & 77.48 & 69.39 & \textbf{84.48} & 47.50 & 76.56 \\
        \hline\hline
\end{tabular}
}
\centering
    \footnotesize{
        \begin{flushleft}
             Note: The superscript $^{E}$ denotes that only the ECN is input into the SVM classifier for evaluating the classification performance. The superscript $^{T}$ denotes that only the TCN is input into the SVM classifier for evaluating the classification performance.
        \end{flushleft}
   }
\end{table*}

\subsection{Experimental Settings}
\label{sub_procedure}

\par For a fair comparison, the same feature selection strategy and classifier are used to test the performance of each brain network model. Specifically, we take the weight of network edges as features, and employ the recursive feature elimination (RFE) \cite{guyon2002gene} strategy to perform feature selection. Finally, a support vector machine (SVM) \cite{cortes1995support} is utilized for classification. The number of selected features by RFE is set to 1000. The parameters of SVM are set as follows: the number of iterations is 1000, the kernel is 'linear', and the regularization parameter is set to 1.0.

\par In this study, three binary classification tasks (i.e., NC vs. eMCI, NC vs. LMCI, and eMCI vs. LMCI) are conducted to evaluate the performance of the proposed method. Prediction accuracy (ACC), sensitivity (SEN), specificity (SPE), and F1 score are used as evaluation metrics. It is worth noting that our method can obtain two kinds of brain network models, including the effective connectivity network (ECN) and the temporal-lag connectivity network (TCN), while other methods can only obtain one kind of brain network model. For a fair comparison, instead of fusing the features of both, we feed ECN and TCN into the classifier separately to evaluate their classification performance. Besides, we conduct a standard 10-fold cross-validation strategy to evaluate the classification performance of all brain network models.

\subsection{Classification Results}
\label{sub_classification}

\par The comparison results of all methods are summarized in Table \ref{classification}, and the best scores are highlighted in bold. From Table \ref{classification}, it can be seen that our method achieves the best performance on all three tasks, which demonstrates the effectiveness of our method. Furthermore, we can find that SLR has the worst performance among the four correlation-based methods, while the classification performance of the other three methods is roughly the same. The reasons for this can be attributed to the following two points. First, SLR introduces two types of constraints at the same time, which makes the constructed brain network too sparse. As a result, it is difficult to extract effective features for the classifier training, so SLR shows the worst classification performance. Secondly, sparsity or low-rank constraints may not be very effective in eliminating the effects of confounding factors. Compared with PC, although SR and LR remove some pseudo-functional connections, the number of effective features retained by the two methods may be roughly the same as the number of effective features extracted from the brain networks constructed by the PC method. Therefore, their classification performance in these three tasks is roughly the same. 

However, the classification results of GC, LiNGAM, and TE are not satisfactory in comparison to correlation-based methods. The reasons for the poor performance of the three methods are different. The reasons for the failure of GC mainly lie in two points: on the one hand, the brain network it constructs is a binary graph, which can obtain fewer effective features than the weight graph; on the other hand, it is because the temporal-lag values between all brain regions are set to a fixed value, which is obviously contrary to the actual situation. Similarly, the reasons for the failure of LiNGAM can also be attributed to two points: one is the addition of acyclic constraint to the brain network, resulting in the constructed brain network being too sparse and not having enough features to train the classifier; the other is that it can only model the linear relationship between brain regions, and cannot fully explore the deeper nonlinear relationship. The reason for the failure of TE is similar to that of GC. The range of temporal-lag values for TE is at least one full TR, which is clearly inconsistent with previous findings \cite{raatikainen2020dynamic, rajna2015detection}.

\subsection{Ablation Study}
\label{sub_ablation}
\par The main contribution of this paper is to design a dual-branch model, which can effectively estimate causal and temporal-lag effects between brain regions. Besides, in order to obtain more accurate estimation results, we also introduce three constraint rules to guide the modeling of the brain network. To verify whether each of these innovative components contributes to the excellent performance, we design four degraded networks in the ablation study, including 1) we remove the temporal-lag inference branch, denoted ``ETLN\_LagGate", 2) we remove the local consistency loss from the loss function of ETLN, denoted ``ETLN\_local", 3) we replace the adaptive mechanism with a fixed threshold of 0.5, denoted ``ETLN\_adaptive", and 4) we remove the spatial constraint loss from the loss function of ETLN, denoted ``ETLN\_spatial".

\par
\begin{figure}[htb]
\centering
\includegraphics[width=0.95\textwidth]{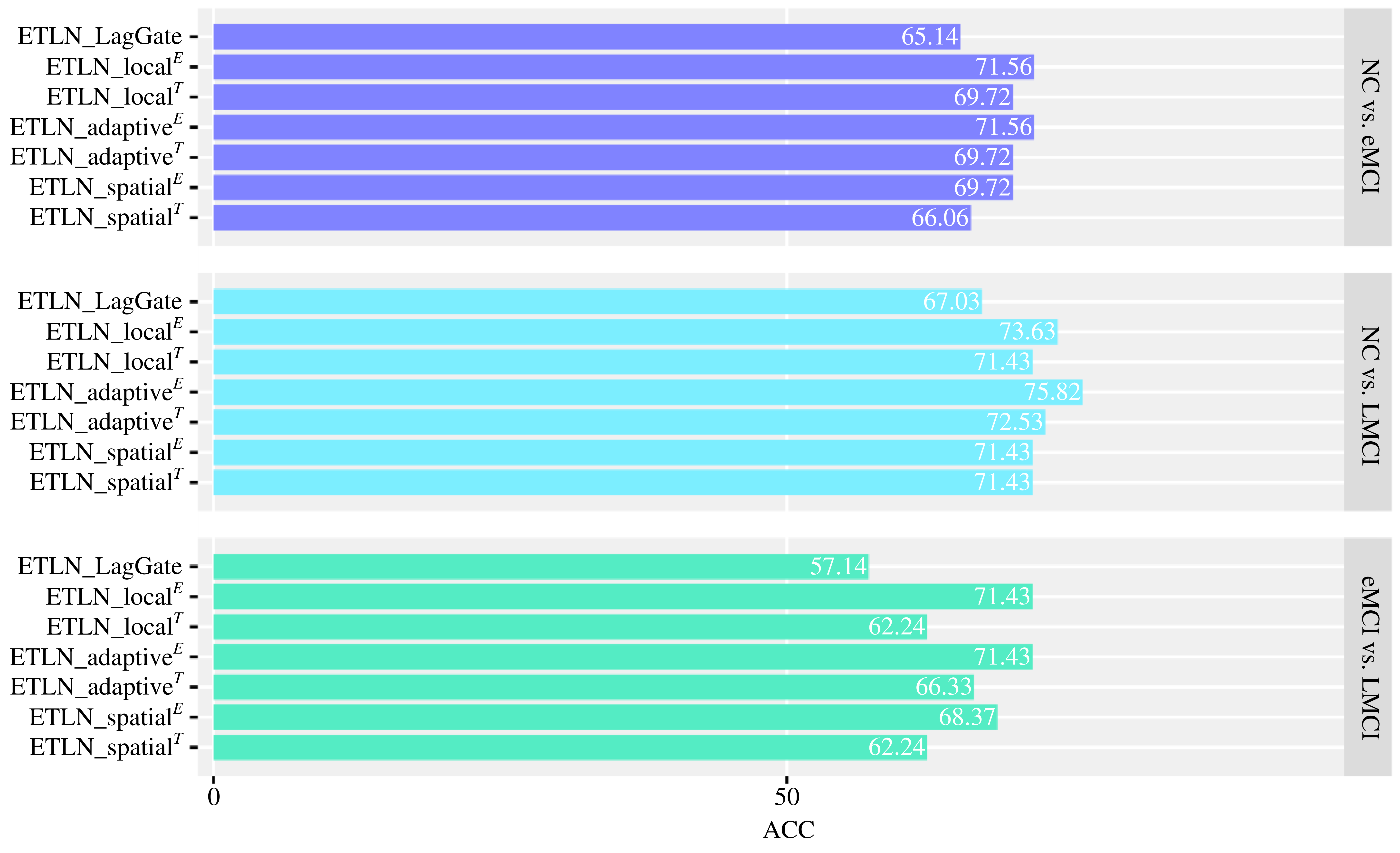}
\caption{Recognition performance for ablation studies.}
\label{ablation}
\end{figure}

 Fig. \ref{ablation} show the experimental results of the ablation study. It can be seen that ETLN\_LagGate obtains the worst performance among all variant methods due to ignoring the effect of temporal-lag. This result suggests that considering the effect of temporal-lag helps to construct better brain network models. Comparing the classification results of the other three variants with ETLN, it can be found that the variant that removes the spatial constraint mechanism has the most performance degradation, followed by the local consistency constraint mechanism, and the adaptive mechanism has the least impact on the result. This result shows that the three constraint mechanisms can be helpful to improve the classification performance, but the spatial constraint mechanism is more helpful, followed by the consistency constraint mechanism, and the adaptive mechanism helps the least.

\section{Discussion}

\subsection{Most Discriminative Patterns}

\begin{figure*}[htb]
\centering
\includegraphics[width=1.0\textwidth]{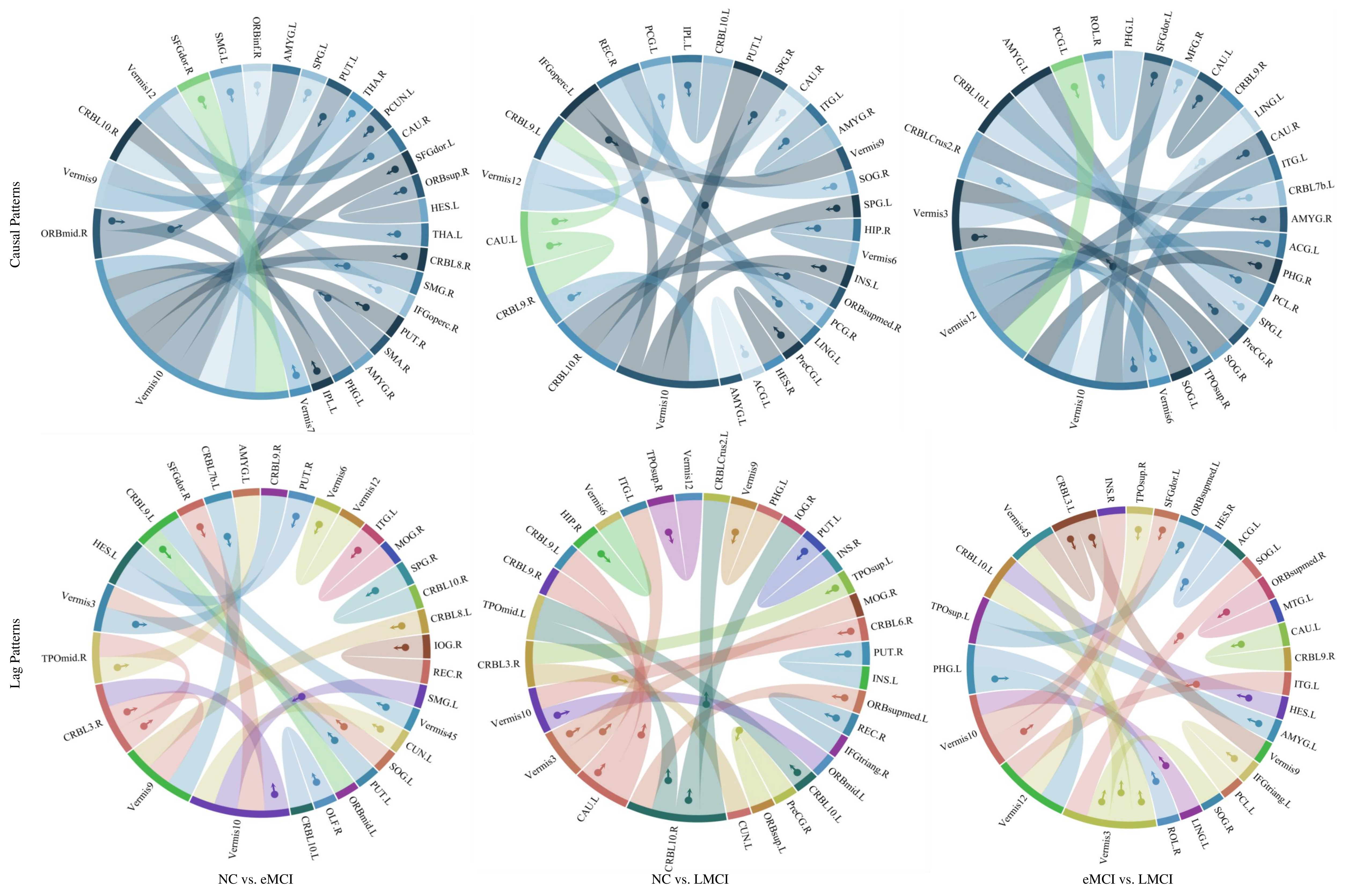}
\caption{Visualization results of the most discriminating causal and lag patterns among the three tasks.}
\label{discriminative patterns}
\end{figure*}

\par Fig. \ref{discriminative patterns} displays the top 20 most discriminative causal and temporal-lag patterns across the three tasks, respectively. There are three points to note about the meaning of the circos graph. First, the color of each arc in the circos graph is randomly assigned. Second, the wider the width of each arc, the more important the corresponding connection. Finally, the motion direction of the ball in each arc represents the causal relationship between two brain regions (from cause to effect).

\par Comparing the most discriminative causal patterns identified by the three tasks, we can draw some interesting observations. First, many brain regions were jointly recognized as potential biomarkers for dementia identification by the three tasks, including the Vermis10, the Vermis12, the left superior parietal gyrus (SPG.L), the right caudate nucleus (CAU.R) and the amygdala (AMYG). This suggests that these brain regions may be potential biomarkers for dementia recognition. Second, all three tasks identified Vermis10 as the brain region containing the most discriminative information. The reason for this may be that this brain region is relatively close to the corpus callosum, and much communication information between brain regions needs to be further transmitted through this brain region. Third, it can be found that the classifier tends to give higher discriminative weights to those functional connections that are spatially distant, especially the connections between the brain and cerebellar regions. We think this phenomenon is reasonable. The transmission of information flow between distant brain regions may need to span multiple brain regions, and abnormal alteration in any node in the transmission path may lead to changes in the strength of functional connectivity.

\par Similarly, comparing the most discriminative lag patterns identified by the three tasks, we can also find some interesting phenomena. Likewise, many brain regions were also jointly identified as biomarkers by the three tasks, including the Vermis10, the Vermis9, the Vermis3 and the left inferior temporal gyrus (ITG.L). But unlike the causal patterns, the most discriminative brain region of the three lag patterns was not consistent. It can also be observed that the classifier prefers to assign higher discriminative weights to those functional connections that are farther away. 

\par Comparing the causal pattern and the lag pattern of the same task, many functional connections are jointly identified as discriminative features. It is clear that these functional connections may play a pivotal role in the determination of classification results. Specifically, for the task NC vs. eMCI, the path from SFGdor.R to Vermis10 and the path from SMG.L to Vermis10 are jointly identified as abnormal connections. For the task NC vs. LMCI, the common abnormal connections include the path from CAU.L to CRBL9.R, the path from CAU.L to CRBL9.L and the path from HIP.R to Vermis6. The reason for this may be due to the introduction of the local consistency mechanism as well as the spatial constraint mechanism. For those brain regions without causality, the local consistency mechanism constrains the temporal-lag values between them to zero. For those brain regions with causality, the spatial constraint mechanism constrains the data distribution of temporal-lag values between them. They both guarantee the local similarity between causal and lag patterns.

\section{Conclusion}
\par In this paper, we propose a novel framework for inferring the causal effects and the temporal-lag values between brain regions. The first point is the design of the network structure, which embeds the target of the solution into the network model as the parameters to be learned. The second point is the introduction of three mechanisms to guide the modeling of brain networks. The whole network is trained in an end-to-end manner and achieved excellent performance on the ADNI dataset. The proposed method not only improves the classification performance, but also provides a new solution for the inference of causal and temporal-lag relationships among large-scale nodes.

\bibliographystyle{plain}
\bibliography{main}

\end{document}